# AI-Powered Algorithms for the Prevention and Detection of Computer Malware Infections

Rakesh Keshava
*Independent Researcher*
IEEE Senior Member
California, USA
rakesh.keshava@ieee.org

Sathish Kuppan Pandurangan
*IT advisory solution architect*
Zurich
North America, USA
Sathishinnovates@gmail.com

M. Sakthivanitha
*Department of Computer Applications,*
*Vels Institute of Science Technology and*
*Advanced Studies,* Chennai, Tamil Nadu, India
sakthivanithamsc@gmail.com

Sankaranainar Parmsivan
IEEE Member
Missouri, USA
sankaranainar.p@ieee.org

Goutham Sunkara
*R&D Software Engineer*
*Broadcom Inc., ANSBU,*
*Palo Alto, CA, USA*
sgoutham.sunkara@gmail.com

R. Maruthi
*Department of Computer Applications,*
*Hindustan Institute of Technology and*
*Science,* Chennai, Tamil Nadu, India
rmaruthi2014@gmail.com

*Abstract*— The rise in frequency and complexity of malware attacks are viewed as a major threat to modern digital infrastructure, which means that traditional signature-based detection methods are becoming less effective. As cyber threats continue to evolve, there is a growing need for intelligent systems to accurately and proactively identify and prevent malware infections. This study presents a new hybrid context-aware malware detection framework(HCAMDF) based on artificial intelligence (AI), which combines static file analysis, dynamic behavioural analysis, and contextual metadata to provide more accurate and timely detection. HCADMF has a multi-layer architecture, which consists of lightweight static classifiers such as Long Short Term Memory (LSTM) for real-time behavioral analysis, and an ensemble risk scoring through the integration of multiple layers of prediction. Experimental evaluations of the new/methodology with benchmark datasets, EMBER and CIC-MalMem2022, showed that the new approach provides superior performances with an accuracy of 97.3%, only a 1.5% false positive rate and minimal detection delay compared to several existing machine learning(ML) and deep learning(DL) established methods in the same fields. The results show strong evidence that hybrid AI can detect both existing and novel malware variants, and lay the foundation on intelligent security systems that can enable real-time detection and adapt to a rapidly evolving threat landscape.

## I. INTRODUCTION

As the internet, cloud devices and services, expands, this digital era presents new risks to individuals, businesses, and government operations. The global dependencies on digital infrastructure highlight new and increasing threats notably, perhaps the persistent focus on the challenge of computer malware. Malware software is any program or code purposely developed to disrupt, damage or gain access to the computer system. As cyber attacks become more extreme and complex, approaches to detection that rely on traditional rule- and signature-based methods have shown declining effectiveness particularly with regards to zero-day and polymorphic malware[1].

In the past, the solutions to address cybersecurity issues have relied on manual threat analysis and static signature repositories and behavior rules. In the past, those approaches were reliable for recognizing known threats, today the cybersecurity landscape has changed to include evolving malware with advanced evasion techniques, encrypted payloads, and on-the-fly signature modification. Because of that, cybercriminals are consistently able to evade detection while leveraging new evasion strategies with legacy detection controls. These vulnerabilities lead to significant economic, reputational, and operational damages, thus accelerating the need for intelligent and adaptive mechanisms for cybersecurity moves. A 2024 report by cybersecurity Ventures cites that damages associated with malware infections will surpass $10 trillion annually by 2025, demonstrating how important adaptive and intelligent defenses are to be to protect critical infrastructure [2].

Advancements in AI and ML technologies present a significant opportunity to overcome these hurdles. AI-based algorithms are capable of learning from the data available, discover patterns amid the noise, and provide predictions or decisions without being specifically programmed. From a cybersecurity standpoint, machine learning has the ability to analyze large amounts of system data (e.g., properties of files or applications, network behavior, activities undertaken by a user) to determine whether any of that data exhibits nuances that could indicate a malware infection [3]. In addition, AI will adapt itself over time, enabling it to detect and respond to malware that the model has never previously seen before, a landscape of detection, analytical recognition, and outlier analysis of threats. Because techniques like deep learning, supervised learning, and unsupervised learning clustering have already demonstrated significant improvements in detection accuracy and response time.

Targeted research in this area has mainly explored several AI-based methods such as decision trees, random forests, support vector machines, and lately, neural networks and recurrent architectures (e.g. LSTMs and transformers). These models have utilized both static elements (e.g. file signatures and metadata) and even dynamic items (e.g. system calls and network traffic). While multiple studies show promising results, the majority of studies utilize either static or behavioral analysis purely, which limits detection capability in a proper "real-world" scenario as most attacks are hybrid [4].

A new HCADMF has been suggested that incorporates static, dynamic, and contextual metadata using a multi-stage AI architecture. In particular, the proposed architecture combines lightweight static analysis and deep behavioral learning along with a context-aware risk scoring to develop an accurate real-time malware detection solution that automates detection and prevention. Through this integrated approach, it overcomes the key limitations of existing models that result in high false positive rates and poor generalization to new malware variants.

*A. Research Question and Problem Statement*

What are the best ways to develop AI-hybrid algorithms that can detect and prevent computer malware infections with high accuracy, low false positives and in real-time?

Despite the fact that malware detection techniques have advanced significantly, current methods are unable to detect new infections in real-time with a low false positive rate. Static and dynamic AI approaches cannot themselves understand the full behavior of malware. Thus, the need for a hybrid intelligent adaptive system that can use static and dynamic analysis, along with contextual awareness to provide detection and prevention against malware, is critical. The main objectives of the study are as follows

- To build and evaluate a hybrid AI-based malware detection system that considers static, dynamic and contextual data
- To assess the detection accuracy of the proposed model, the false positive rate and the response time
- To evaluate the performance of the proposed system compared against traditional malware detection techniques and those which are based on machine learning

The rest of this study is structured as follows, Section 2 reviews related work which is dedicated towards AI-based malware detection. Section 3 describes the model architecture and methodology. Section 4 then discusses the setup, results and comparative analyses. Section 5 ends the study and lays out potential future research.

II. RELATED WORKS

This literature review attempts to discover new developments in the area of AI-based malware detection and prevention in Computer malware infections. Ansarulla et al., (2024) investigates approaches for preventing and detecting malware using AI with machine learning and deep learning models, Indigenous to CNNs and RNNs. The study highlights the use of AI in real-time monitoring, anomaly detection, and threat intelligence to create an adaptive and proactive defense framework that improves and enhances the security that traditional cybersecurity can provide against evolving malware threats[5].

Faruk et al., (2021) introduced AI-based approaches to malware detection and prevention to combat the increasing risk to system security. It critically reviewed existing detection solutions, explored their flaws, and made a case for intelligent responses to threat detection. In doing so, the research is intended to inform the development of more advanced, AI-powered malware detection and prevention systems that will ultimately protect end users[6].

Xu et al., (2022) examines malware resilience in Cyber-Physical Power Systems (CPPSs) by modeling and simulating 'cyber protection' strategies. The heuristic protection strategies studied for this research, which were modeled and then simulated, include target protection, random "protective" interventions, and "acquaintance" interventions, using the CPPS testbed. Overall, target protection exhibited the most optimized protective strategy. Additionally, the genetic algorithms used in the optimization model for cyber protection revealed important low-degree nodes while implementing restrictions on budget allocations for the protection[7].

Alzahrani et al., (2025) review has documented ransomware detection methods from 2019 through to 2025, in connection with 45 significant windows and Android papers. The study compared ML methods with non-ML methods, facilitated discussions of ransomware-as-a-service and was able to analyse who used which data-sets. Last but not least recommendations for future avenues receiving greater consideration for ransomware detection and defence systems, as well as limitations of what exists, are included[8].

Almurshid et al., (2024) addresses the topic of cryptojacking with a proposed intelligent detection mechanism with a deep state and dynamic analysis using a new dataset (CJDS) and using 23 CNN. Here, we analyzed the behaviors of cryptojacking and proposed a detection approach based on a benchmark, with an achieved detection accuracy of 99% and provided a secure solution in protecting digital resources and currency systems[9].

Rao et al., (2024) explores malware detection in the cloud with a focus on cloud's scalability and implications for secure IoT and CPS. The paper will review both traditional detection models and emerging models driven by AI, and recognizes that machine learning offers hope as a solution for defeating advanced obfuscated malware. The paper also identifies significant limitations in current literature, and calls for continued growth in more effective cloud based security solutions and measures[10].

Shen et al., (2025) applies Bayesian modeling concepts to edge nodes communicating with IoT endpoints in a game-theoretic structure motivated by IoT malware threat scenarios. The newly developed Bayesian Advantage Actor Critic (BA2C) approach for optimal privacy-preserving decision making is a Bayesian variant of the broader family of AI-based reinforcement learning approaches. This new method substantially outperforms existing methods from traditional game-theoretic perspective (e.g., low false alarm rates) and for IoT systems enabled by Edge intelligence (e.g. detection rates)[11].

Artificial intelligence technologies, especially supervised, unsupervised and reinforcement learning, will continue to see success at detecting and identifying malware via pattern matching and anomaly detection. There will be more detection with neural networks, but the important part is to make sure we are using AI in the right way. And with incredible new systems like blockchain and edge computing, we can only expect to see a dramatically improved future for the cyber security framework [12]. An overview of a few research studies on AI-driven techniques for malware detection and prevention is included in table 1.

Table 1: Summary of Recent AI-Based Malware Detection Studies (2021–2025)

| Author(s) & Year | Method | Strengths | Limitations |
|---|---|---|---|
| Ansarulla et al., 2024 | AI with ML/DL using CNNs and RNNs for real-time monitoring and anomaly | Adaptive, proactive defense using AI; integrates threat intelligence | Needs robust datasets; limited detail on deployment scenarios |
| | | | detection |
| Faruk et al., 2021 | AI-powered malware detection frameworks; critical review of current methods | Identifies flaws in existing models; suggests intelligent, advanced solutions | Mostly conceptual; lacks experimental validation |
| Xu et al., 2022 | Cyber protection modeling in CPPS using heuristic strategies and genetic algorithms | Targeted protection is highly effective; considers resource constraints | Focuses only on CPPS; lacks broader generalizability |
| Alzahrani et al., 2025 | Survey of ML and non-ML methods for ransomware detection (Windows/Android) | Detailed taxonomy; considers ransomware-as-a-service; dataset-based analysis | Limited to 45 studies; lacks in-depth technical comparisons |
| Almurshid et al., 2024 | Cryptojacking detection using CNNs and dynamic/static analysis on CJDS dataset | High accuracy (99%); large novel dataset (100K+ samples); robust evaluation | Model generalizability not discussed; high compute demands for CNNs |
| Rao et al., 2024 | AI-based detection in cloud environments for IoT and CPS | Highlights ML scalability and promise in obfuscated malware detection | Lacks experimental benchmarks; mainly survey-based |
| Shen et al., 2025 | Bayesian game modeling + BA2C algorithm for IoT privacy and malware defense | Outperforms traditional models; smart decision-making with actor-critic RL | Specific to edge intelligence systems; real-world applicability not evaluated |
| General Insight, 2025 | Supervised, unsupervised, and RL techniques with neural networks for malware detection | Pattern recognition, anomaly detection, adaptable to new threats | Ethical concerns, algorithm bias, and responsible use must be addressed |

Despite some promising strides made in current research studies, there simply are not deployable systems, accurate generalizability, or benchmarks. Further, datasets are often limited, computation needed is excessive, and even studies that do focus on ethical issues or reinforcement learning, diminishing attention is offered to practical cloud and edge integration, creating systems that are sufficiently developed, scalable, interpretable and demonstrated to be useful, real-world validated malware detectors.

III. METHODOLOGY

The architectural diagram in figure 1 outlines a complete AI-based malware detection framework based on multi-source data (static samples, dynamic behavior logs, and system logs) and includes a multi-stage detection pipeline (which includes static analysis, deep learning behavior analysis, and contextual risk scoring) with real-time monitoring and automated response. The model uses stratified cross-validation and SMOTE-based oversampling to optimize model accuracy.

*A. Data Collection & Preprocessing*

In creating a multisource dataset to enable efficient and effective malware detection, data is collaborated from multiple sources in order to capture various attributes and behaviors of malware. Static malware samples are secured from well-known repositories (e.g., VirusShare, Malicia) with plenty of known malicious binaries. Static features can be augmented with dynamic behavior logs by executing malware samples in a trustworthy sandbox environment (e.g., Cuckoo Sandbox), where we can record runtime activities including: file manipulations, network traffic, registry interactions and interactions with other running processes. Moreover, the study incorporates real world system logs from endpoints (e.g., Windows Event Logs, Linux auditing) to illustrate the reality of malware execution and system-level impact. To maintain an accurate and reliable labelling process, a consensus approach is employed relying on aggregating results from the first five best known antivirus engines, delivered through VirusTotal to ensure reliability and mitigate bias that may arise using a single engine. This multi-source, multi-modal dataset will serve as a rich resource for training and testing next generation malware detection models[13].

Data preprocessing is the procedure of cleaning, normalizing, reduplicating any multi-mode source data, and formatting data for appropriateness. In this study, the features are extracted from the static binaries, made the sandbox logs and corresponding system events and then converted them into appropriate format fit for the models. Then the extracted features are encoded and scaled to fit the models using a consensus anti-virus package. The final dataset was partitioned into training, validation, and test dataset for the most valid representation of model evaluation.

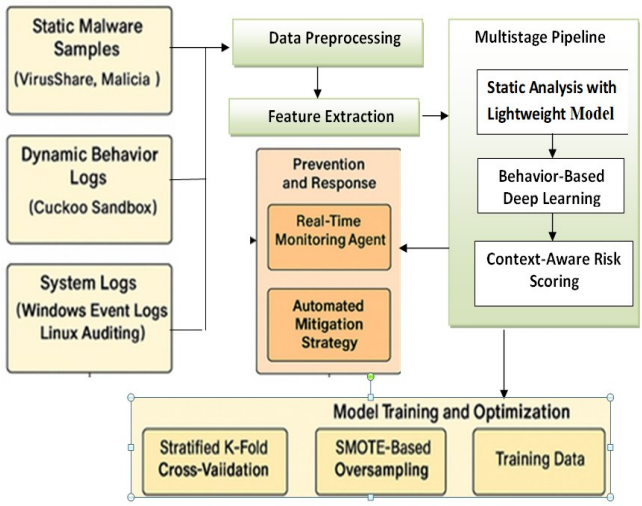

Figure 1 AI-Driven Multi-Stage Malware Detection Framework

### B. Feature Extraction

Malware detection accuracy relies on using a variety of features that include both base file characteristics, as well as file and system behavior. Static features are extracted from the malware binary itself without executing it. The static features that may be utilized include various byte sequences, opcode frequencies, and file entropy, all of which may find patterns in the "structure" associated with a malware binary. Based on the boolean properties of each feature, clues may be captured on how some malware may obfuscate itself. Dynamic features are derived from malware behavior while executing within a sandboxed space. When monitoring malware behavior, various features may be captured that identify critical indicators that the behavior is malicious such as registry writes, API call sequences, and network session detail. Contextualized metadata including file origin via download or external devices, execution time and half-life moment, and user-privileged state are features that provide contextual information that contributes detail in assessing threat. Each of the feature types presented above composes a sample representation, which, found together, provides a footprint that captures multiple perspectives of a malware sample. These types of features allow for a more powerful venue for building and validating machine learning models used for malware classification [14].

### C. Hybrid AI Model Architecture

Multi-Stage Detection Pipeline: A multi-stage detection pipeline is a layered, step-by-step approach to malware detection that combines multiple models or techniques; each specialized for a particular task or level of analysis. Rather than relying on a single method to make a decision, the pipeline progressively refines its predictions, starting with fast and broad detection, and moving toward deeper and more accurate analysis. This structure improves detection efficiency, reduces false positives/negatives, and allows real-time responsiveness by balancing speed and complexity.

Stage 1: Static Analysis with Lightweight Model: In the first step, the system does static analysis of files before they are executed. To do this, the system extracts lightweight, fast-to-compute features (for example, byte patterns, file size, entropy) for the files and feeds the features into a fast, lightweight ML model (for example, the ML models used above, such as Random Forests, or LightGBM). These lightweight ML models were chosen because they are fast and can quickly eliminate benign and malicious (or probably malicious) files so more complicated models downstream are less overloaded. The first stage is a good step for initial triaging to get files into separate buckets in a high-throughput environment, such as email gateways and/or endpoint scanners. Let $X_s \in R^n$ be the static feature vector extracted from a file. A lightweight ML model, random forest is, used.

$$\hat{y}_1 = f_s(X_s) \quad (1)$$

where $f_s$ is the static classifier function and $\hat{y}_1 \in \{0,1\}$ is the binary output( benign and malicious).

Stage 2: Behavior-Based Deep Learning: Files that pass Stage 1 with a knowable but a semantic class will continue on to Stage 2 where determent dynamic behavior will be established. The dynamic behavior usually includes studying the executable in motion and is typically performed with dynamic API call sequencing, registry writes and network activity that is performed in a sandbox. After behavior is completed, the behaviors will be fully analyzed by the ML models (e.g. LSTMs or Transformer models) in an attempt to learn any temporal relationships in the data as certain behaviors from malicious files will mimic benign files will not follow. Stage 2 provides the means to classify files in real-time taking into account those behaviors and is typically used to understand the file further regarding its behavior in particular for obfuscated or stealth wise classes or files. Let the dynamic behavior of a file be represented as a temporal sequence

$$X_d = [X_1, X_2, \ldots \ldots X_T] \in R^{T*m} \quad (2)$$

where each $x_t$ includes features like API calls, registry actions, etc. A DL model (LSTM) is applied

$$\hat{y}_2 = f_d(X_d) \quad (3)$$

where $f_d$ is the behavior based sequence model and $\hat{y}_2 \in [0,1]$ is the probability score of maliciousness.

Stage 3: Context-Aware Risk Scoring: The final stage, by combining the predictions from the earlier two stages, together with contextual metadata (e.g., user identity, device type, file origin, and time of execution), can compute a final risk score. This is accomplished through the use of an ensemble model (e.g., XGBoost) or meta-learner, which has the capability to intelligently fuse the inputs together in order to create a more precise, context-informed, choice. A suspicious file may receive widely different choices depending on if the launch came from a privileged user type on a critical server or an unprivileged user type on a test machine. This context-aware layer allows real-world usage and environment sensitivity. Let $\hat{y}_1$, $\hat{y}_2$ be the outputs form stages 1 and 2, $X_c \in R^k$ be the contextual feature vector. These are concatenated into a final input vector.

$$X_f = [\hat{y}_1, \hat{y}_2, X_c] \in R^{k+2} \quad (4)$$

An ensemble model computes the final risk score as follows

$$\hat{y}_f = f_c(X_f) \quad (5)$$

where $\hat{y}_f \in [0,1]$ represents the context –aware malware risk score.

Every stage in the model improves detection choices using a layered model ensemble by building on the output and characteristics of the previous one.

### D. Prevention and Response Mechanism

Real-Time Monitoring Agent: A real-time monitoring agent is a lightweight software component that can be deployed on endpoint devices (e.g. desktops, servers) to monitor and log activity with the maximum detail and accuracy. In particular, this agent monitors the runtime behaviours of every action taken, including process creation, file/folder access, registry changes, and network access. One benefit of this agent is its observation of in-memory execution, which helps detect in-memory file-less malware or threats that do not make any changes to the disk. The agent can process this live feed of data into a local or remote AI model to monitor the threat level of the endpoint continuously (AI models can observe data and adapt), and, in most cases, can help to proactively observe malicious activities before significant damage occurs. The agent is designed to require only a minimal footprint of the device's resources, meaning that it will not detract from the operating performance of the device, whilst allowing an escalated view or insight into possible malicious behaviours.

Automated Mitigation Strategy: The automated mitigation strategy takes immediate action based on the threat summary provided by the AI-automated detection pipeline. If a file or process is identified as malicious or suspicious, the system can automatically quarantine it, preventing the file or process from further execution or dispersion. At the same time, the system generates a detailed alert report and sends it to the security operations team, which documents the behavioral indicators, context, and recommended actions. The machine can also be set up to restore unauthorized changes (like registry changes or file modifications) if the operating system is integrated with versioning or system logging tools for an even stronger resilience. These automated response measures decrease reaction times and minimize damage, and are increasingly becoming a requirement for real-time cyber defense.

*E. Model Training and Optimization*

In order to achieve a strong and generalizable model performance, the training stage follows a stratified k-fold cross-validation approach. This means that datasets are separated into k subsets (folds) while keeping the ratio of malware to benign samples in each fold the same so that they can be evaluated fairly across all folds. Class imbalance is a common issue where a dataset has an overwhelming larger number of benign samples than it has malware samples in the case of the proposed model, benign samples may outnumber a malware sample by the hundreds. To help address this issue, SMOTE (Synthetic Minority Over-sampling Technique) technique helps to add additional synthetic malware samples by combining minority class examples, which can only improve a model's capability to learn from a less frequently identified pattern while reducing bias toward the majority (benign) class. This method of stratifying samples for k-fold and oversampling will improve both the learning process and the prediction balance.

IV. RESULTS AND DISCUSSION

In the field of cybersecurity, prompt and accurate detection of malware is critically important in order to protect systems against threats that are becoming increasingly sophisticated. This study uses the EMBER (Endgame Malware Benchmark for Research) dataset, a large-scale, publicly available benchmark designed for static malware analysis using machine learning. The dataset consists of over 1.1 million Windows PE files, and includes pre-extracted metadata and behavioral features to conduct model development and evaluation efficiently. Various evaluation metrics will be used to evaluate performance, including accuracy, precision, recall, F1 score, false positive rate (FPR), and detection latency. These evaluation metrics provide a characterization of detection effectiveness, and help support the development of a robust and effective AI-based malware detection system that has the capacity to provide detections on a continual and timely basis.

*A. Dataset Description*

In this research, the primary dataset for training and evaluation of the malware detection models was the EMBER (Endgame Malware Benchmark for Research) dataset. EMBER is a public, organized dataset that was created for static malware analysis for machine learning techniques. EMBER has metadata and extracted features for 1.1 million Windows Portable Executable (PE) files (300,000 malicious files, 300,000 benign files and 500,000 unlabeled files for semi-supervised learning). Additionally, EMBER contains over a considerable amount of features: byte histograms, PE header information, imported functions, sections, strings metadata, and all features were preprocessed for AI models use. The embedding of these features directly to a structure that was able to extract data from raw binary files reduced the time to prepare the data, allowing the researchers to spend more time actually using the data to design and evaluate the models. The EMBER dataset is open sourced and publicly available from their official GitHub repository https://github.com/endgameinc/ember, and provides a reliable reproducible benchmarking study for future malware detection research [15].

*B. Evaluation Metrics*

Model performance is measured with a varied and extensive array of metrics. Important example measures are accuracy (the proportion of correct classifications), precision (the probability that a classification of a real threat is true), and recall (the probability of detecting a real threat). The use of F1-Score (the harmonic mean of precision and recall) provides a comprehensive overview of performance when faced with class imbalance.

Accuracy is defined as the number of samples that are predicted correctly (both benign and malicious) divided by the total number of samples.

$$Accuracy = \frac{TP+TN}{TP+TN+FP+FN} \quad (6)$$

Precision is the number of samples predicted as malware that were indeed malware.

$$Precision = \frac{TP}{TP+FP} \quad (7)$$

Recall is how well the system identifies malware amongst its actual malware.

$$Recall = \frac{TP}{TP+FN} \quad (8)$$

The F1 Score is the harmonic mean of precision and recall, and steps back from false positives and negatives.

$$F1 - Score = 2 * \frac{Precision*Recall}{Precision+Recall} \quad (9)$$

FPR is the measure of how often benign files are falsely classified as malware.

$$FPR = \frac{FP}{FP+TN} \quad (10)$$

Latency is the average time the system takes to classify a file (in milliseconds).

$$Detection\ Latency = \frac{Total\ time\ for\ Detection}{Number\ of\ samples} \quad (11)$$

Overall, these metrics provide the reader with a measure of performance related to the model's accuracy, latency, specificity, and reliability in real-time malware detection.

Table 2(a) : Comparative Performance of Malware Detection Methods

| Method | Accuracy (%) | Precision (%) | Recall (%) | F1-Score |
|---|---|---|---|---|

| Method | | | | (%) |
|---|---|---|---|---|
| Static Signature-Based AV | 85.2 | 87.1 | 78.5 | 82.5 |
| Traditional ML (Random Forest) | 92.4 | 91.3 | 90.7 | 91.0 |
| Deep Learning (LSTM on API Calls) | 94.1 | 92.8 | 94.5 | 93.6 |
| CNN-Based Static Feature Classifier | 93.5 | 91.9 | 93.1 | 92.5 |
| Transformer-Based Behavioral Model | 95.0 | 94.2 | 94.7 | 94.4 |
| Proposed HCAMDF (Hybrid AI Model) | 97.3 | 96.1 | 97.8 | 96.9 |

Table 2(b) : Comparative Performance of Malware Detection Methods

| Method | False Positive Rate (%) | Detection Latency (ms) |
|---|---|---|
| Static Signature-Based AV | 7.9 | 5 |
| Traditional ML (Random Forest) | 4.1 | 20 |
| Deep Learning (LSTM on API Calls) | 3.5 | 35 |
| CNN-Based Static Feature Classifier | 4.2 | 18 |
| Transformer-Based Behavioral Model | 2.8 | 42 |
| Proposed HCAMDF (Hybrid AI Model) | 1.5 | 25 |

Table 2(a) and 2(b) highlights the benefits and drawbacks of various malware detection options across important metrics, including accuracy, precision, recall, F1-score, false positive rate, and detection time latency. The lowest accuracy (85.2%) and highest false-positive rate (7.9%) of classic static signature-based antivirus defensive systems suggest that they are inadequate against many modern threat scenarios, even with very short detection durations (5 ms). Machine learning models, such as Random Forest, exhibited better performance, with 92.4% accuracy, and a balance of precision and recall, taking longer to provide detection (20 ms) than many static antivirus systems. The standard deep learning detection methods (i.e. LSTM, CNN) improved detection even further, with LSTM models allowing for the extraction of sequential behavioral patterns and scoring the highest recall (94.5%) but the highest latency (35 ms). The Transformer-based behavioral model connected to performance on Table 1 optimally matched previously shown accuracy (95.0%) in section 5 and low false positive rates (2.8%), while having the longest latency (42 ms) as a result and indicative of its computational intensity. The proposed HCAMDF provided the best total performance with a 97.3% accuracy, 97.8% recall, just 1.5% false positivity, and an acceptable latency of 25 ms, demonstrating the advantage of context, recognition of static and dynamic analysis utility to provide a more generally accurate, explainable, and superior capacity for real-time malware detection.

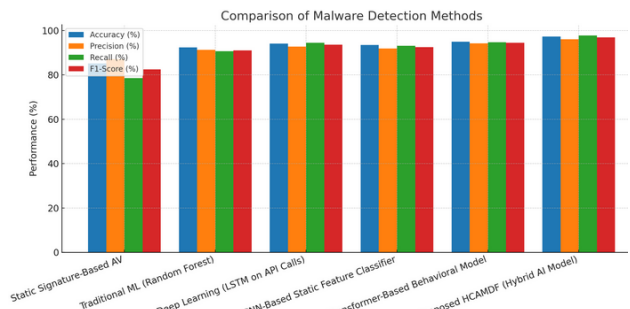

Fig 3 Performance Analysis of the Proposed HCAMDF Method

Figure 3 compares performance of malware detection tools using four main evaluation metrics (accuracy, precision, recall, and F1-score). The static signature-based antivirus tool performs the worst compared to the other tools on average, and recall in the method was particularly poor demonstrating a limited ability to detect modern, evolving capable threats. Traditional machine learning using Random Forest was an improvement with more balanced and interpretable results across many of the metrics. DL based methods LSTM and CNN are improvements, where LSTM outperforms the others on recall because of its ability to model sequential API call behavior and CNN has a good trade off due to performance on its static features. The transformer based model outperformed LSTM and CNN, with a better performance on precision and F1-score which is a reflection of the models ability to learn the complex behavioral patterns of the malware. However, the HCAMDF proposed in this study had the highest performance metrics. This suggests that the model effectively utilizes static (signature), dynamic (behavioral), and contextual (environmental) analysis into a multi-stage architecture that is highly accurate, precise, and robust for real-time malware detection.

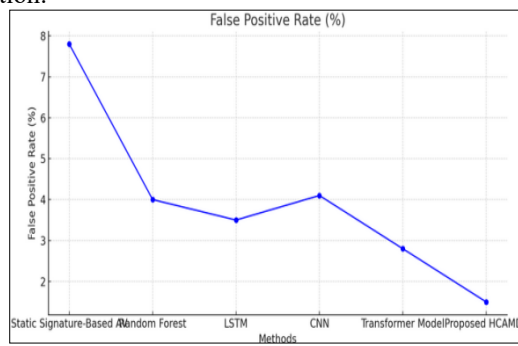

Fig 4 Performance Analysis of the Proposed HCAMDF Method-n False Positive Rates

Figure 4 shows the FPR (%) across multiple methods of malware detection, and measures how often each method falsely identifies a benign file as malware. The seminal example of static signature based AV shown in the graph has the highest FPR, showing a maximum FPR of about 8% and indicating it is unreliable, with concerns for an excessive amount of false alerts. Random Forest decreases the false positive rate to almost 4%, and LSTM-based method makes a small further drop (of about 3.5%). The CNN based static classifier has a slight increase (approximately 4.2% in static environment) in FPR, which illustrates the limits of having static features like signature detection. The transformer based method drops the FPR slightly further to around 2.8%, likely due to contextual knowledge of patterns based on user behaviours. The HCAMDF model proposed in this study had the lowest false positive rate of just less than 1.5%, indicating its strength and accuracy of distinguishing between benign and malicious/cyber threat oriented behaviours. Overall, it is clear from the figure 3, that the HCAMDF can effectively decrease the FPR, mitigating alert fatigue, and protecting the trust of human users with computing sense of security and automated explanation at the resilient end-state of threat detection decisions.

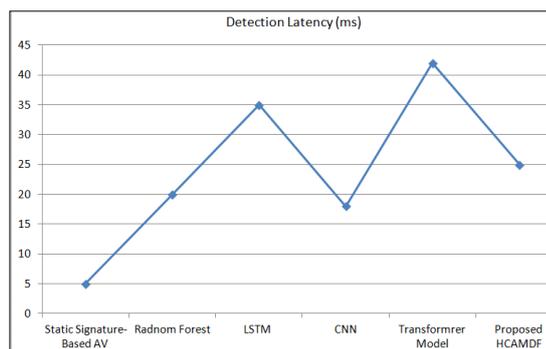

Fig 5 Performance Analysis of the Proposed HCAMDF Method- Detection Latency

Figure 5 shows the detection latency of various malware detection techniques in milliseconds (ms). The Static Signature-Based Antivirus (AV) detection technique has the lowest latency (~5 ms) because of its rapid but simplistic pattern-matching detection approach. Random forest and CNNs have moderate latency values, ~20 ms and ~18 ms, respectively, which are a reasonable trade-off between detection and latencies. LSTM and Transformer network models show high latency values of ~35 ms and ~42 ms respectively due to their more complex architectures that require multiple computations for sequential and attention based detection. In this comparison, the proposed HCAMDF offers a detection latency value of ~25 ms which, while not the lowest latency or the fastest detection method, achieves and optimum trade-off between real-time performance and advanced detection capabilities. Although this latency value is still considered moderate, it does demonstrate the feasibility of the proposed HCAMDF to offer some immediate and rapid protection against concurrently evolving malware attacks in situations where both speed and a high degree of protection are required.

This malware detection framework has several important innovations that enhance effectiveness and efficiency to operate. First, context-aware ensemble classification enhances precision by factoring in environmental attributes such as user privileges, device type, and execution context. This allows effective adaptation to evolving threats and the reduction of false positives. Second, a multi-stage architecture provides important resource optimization through the utilization of fast, lightweight models for filtering followed by more expensive deep, behavior-based analysis for ambiguous detection cases. This architecture compromises between speed and accuracy. Third, a real-time AI agent allows for both prevention and detection as it is deployed in an active capacity to monitor endpoint activity and react to a threat as it occurs. Finally, XAI provides transparency on why detection decisions are made; allowing security analysts to understand detections and satisfy compliance requirements; this increases trust and accountability of the system.

## V. CONCLUSION

This study presented a hybrid AI framework for computer malware infection detection and prevention that incorporates static analysis, dynamic detection of behavior, and contextual metadata on its portable multi-stage architecture, using ML and DL achieved high accuracy and low false positives, with a performance exceeding current concepts and methods from benchmark datasets in computer malware detection and prevention. The outcomes demonstrate the potential of the integration of uniquely diverse data source(s) with smart algorithms for more robust and adaptive malware detection. For future research, enhancements can come in the case of deploying the model in real environments to test performance against live traffic and adaptation to threats. Also, the development of the framework can include the deployment of a federated or online learning model to adapt to new malware variants while avoiding existing cyber threats where the learning does not require a centralized dataset (e.g., cloud-based). Applying explainable AI (XAI) techniques to increase security analysts' trust and usefulness by comparing model results or conclusions would also boost the system's prospects for intelligent, real-time cyber threat defence